# Enhancing Medical Anatomy Education through Virtual Reality (VR): Design, Development, and Evaluation

Myint Zu Than[1], and Kian Meng Yap[2]
[1] Department of Computing and Information Systems, School of Engineering and Technology,
Sunway University, Malaysia
[2] Research Centre For Human-Machine Collaboration (HUMAC), School of Engineering and Technology,
Sunway University, Malaysia
(Email: myintzuthan00@gmail.com;  kmyap@sunway.edu.my)

**Abstract ---** Modern medicine demands innovations in medical education, particularly in the learning of human anatomy, traditionally taught through textbooks, dissections, and lectures. Virtual Reality (VR) has emerged as a promising tool to address the limitations of these conventional methods by emphasising vision-based and active learning. However, current VR educational tools are often inaccessible due to high costs and specialised equipment requirements. This paper details the design and development of an accessible, desktop-based VR system aimed at enhancing anatomy education by leveraging the user's visual perception to promote a meaningful and interactive learning experience. The Virtual Anatomy Lab was designed to enable students to interact with a 3D Skull model to complete tasks virtually via an interactive user interface (UI) with the help of common devices like a mouse and keyboard. As part of the study, a group of medical students from prestigious medical schools throughout Malaysia were invited to evaluate the built system to offer feedback and determine its overall efficiency and usability in fulfilling their learning goals. The results and findings from user evaluations were then analysed to discuss its effectiveness and areas for future improvement.

Keywords: Virtual Reality, Medical Education, Visualisation in Education, Interactive Learning

## 1. INTRODUCTION

Virtual Reality (VR) is an immersive technology which utilises a variety of computer graphic systems to transport the user's cognitive state of mind into an interactive virtual environment where multi-sensory experiences are obtained from engaging with objects virtually. Entities encountered in these spaces often consist of 2-Dimensional (2D) or 3D objects which can be picked up, moved about, heard, smelt, touched, and studied in various perspectives, allowing users to interact with items in VR in a direct and natural manner, much as they would with objects in real life [1]. In the field of medicine, VR has especially emerged as a promising tool for education and training where virtual simulators are used to assist with surgical training, clinical practices, and so on [2]. This is because VR employs virtual learning based on a multidimensional space to provide students with the most abstract and effective methods of acquiring new information and practices. Thus, this technology shows utmost potential to address numerous issues and limitations identified in conventional teaching methods used in medicine, specifically in the study of gross anatomy [3]–[5].

Anatomy forms the cornerstone of medical education, providing essential knowledge of human body structures that are crucial for aspiring medical professionals to practise within the field. However, research indicates that traditional teaching methods, which typically involve textbooks, surgical videos, cadaver dissections, and simulations with dummy models, are often ineffective in aiding students' comprehension of the subject [4]–[8]. The challenge lies in the depiction of unclear 2D images in textbooks, including the high costs and logistical challenges of using human cadavers [9]. Safety concerns related to infections further restrict direct student interaction with cadavers [10, 11], necessitating costly safety measures. Such limitations not only impact educators and faculty staffs, but it primarily effects the students' learning experience. As a result, the lack of fundamental knowledge on such topics can cause students to experience major difficulties during viva examinations that are often conducted to assess their understanding on anatomical structures of the human body [10].

VR offers a promising solution to these challenges, providing immersive environments that enhance learning experiences without the limitations of traditional methods. That said, current VR systems are often prohibitively expensive, posing a barrier to widespread adoption, especially for institutions and individuals with limited resources [11]. The lack of affordability and accessibility of current VR anatomy systems in the market remains a primary obstacle, preventing many users from fully benefiting from its potential educational advantages.

This paper presents the development of a desktop-based VR system designed to enhance medical education by offering an engaging virtual environment for users to explore the five fundamental perspectives of the human skull. It aims to promote active participation for effective learning while addressing the limitations of traditional learning methods used in medicine and fostering accessible educational innovation without requiring high-tech gadgets like

Oculus HMD headsets. Since vision and observation are crucial in anatomy education, the development of the VR system focuses on leveraging this technology to enhance cognitive functions through visual perception, designed to enable users to interact with 3D components to perform virtual tasks.

Section 2 of this paper further outlines the design and development approaches involved in achieving the project's objectives and outcomes. Section 3 of this paper discusses the findings and results from testing activities, comparing them with earlier research in the field. Finally, the conclusion in Section 4 reviews and evaluates the overall work that has been accomplished, presenting insights and ideas for future work and research for the field.

## 2. APPROACHES TO DESIGN AND DEVELOPMENT

The VR system in this study was built on the Unity Engine software platform; a popular choice for VR applications since it enables the building of realistic virtual environments using C# as its core programming language. Earlier studies [3, 4, 9] revealed that the subject of anatomy typically relies on VR technologies to provide high quality and accuracy of images and objects. Hence, the Universal Render Pipeline (URP) template in Unity was chosen for the development of this VR system. The URP template, as opposed to other Unity project templates, optimises the graphical quality of elements included as part of the virtual environment, thus resulting in better experiences for users to visualise 3D objects or imagery in their surroundings.

### 2.1. Construction of the 3D Virtual Anatomy Lab

To address the limitations of traditional anatomy education, a Virtual Anatomy Lab was developed to facilitates self-directed learning, free from ethical constraints, equipment shortages, or the need for constant educator supervision. This virtual environment aims to replicate the experience of a real anatomy lab, featuring realistic 3D entities for users to observe and interact within it. To ensure its accessibility, it was designed for use with readily available personal devices, eliminating the need for specialised equipment. This was accomplished using Unity's XR Interaction Toolkit Package, enabling 3D and 2D graphical interactions via specified mouse and keyboard inputs.

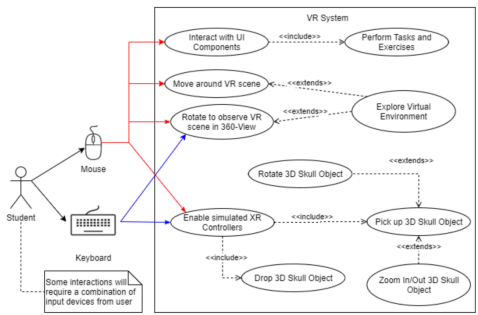

1. Use-Case Diagram of the VR System

An overview of the interaction flows implemented in the VR system is depicted through the use-case diagram above. According to a recent study by Falah et al. [4], the authors believed that the ability for medical students to directly manipulate the relative sizes of 3D organs in the virtual space is a useful interaction in VR since these activities would be difficult to do so in a real life setting. This claim is supported by a similar investigation by Hussein and Natterdal [10] which states that VR enables medical students to perform repetitive tasks and continuous actions without the physical risks involved. With these findings in mind, the Virtual Anatomy Lab is designed with the objective of inducing active learning among students to ensure that they are able to engage with realistic 3D organs and graphical components presented in the VR system repetitively and at their own pace, rather than memorising information off the textbook or video lectures. This fosters experiential and active learning for users as it requires them to actively interact with elements placed inside the virtual world in order to accomplish tasks [11].

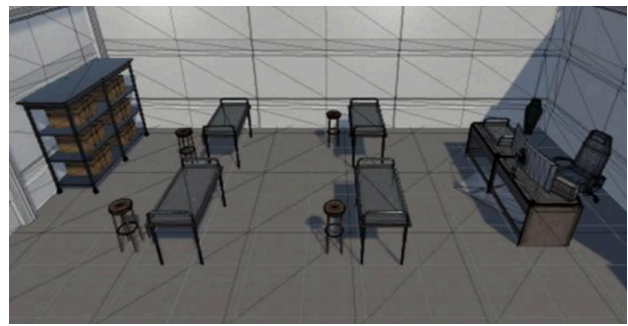

2. Scene of 3D Assets in the Virtual Anatomy Lab

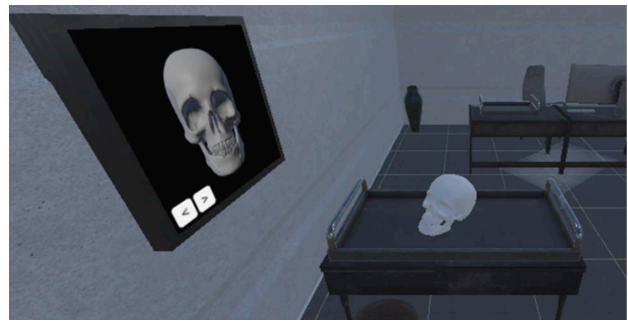

3. Scene of the Student's Dissection Table Area

The figures above illustrate the process of constructing the Virtual Anatomy Lab, which was modelled based on photographs taken of actual anatomy labs from medical institutions. 3D assets, such as the furniture depicted in Figures 2 and 3, were obtained from various sources and enhanced with textured surfaces (e.g. metal trays, wooden stools) to improve their visual appeal and realism. It includes an anatomical accurate and realistic 3D Skull Model as the primary object of the study. It was noted, based on prior investigations done on VR systems in medical education, that users often feel overwhelmed when their virtual environments are cluttered with numerous hyper-realistic objects that may not be relevant to the system's purpose [5]–[11]. Thus, the design and structure of assets in the Virtual Anatomy Lab have been intentionally kept minimal to prevent unwanted distractions and overstimulation of the user's visual senses. This design restriction helps to maintain the user's focus on the primary object of study while minimising distractions from possible interactions with other objects in the same virtual environment. The creation of this Virtual Anatomy Lab is pivotal in delivering an immersive learning experience by captivating the user's visual senses, thereby creating a realistic 3D setting that feels familiar to enhances engagement and learning.

### 2.2. Implementation of Interactions and Tasks

A study by Chavez and Bayonna [11] observed that the "movement" characteristics – which can be defined as an action in altering something's position or speed – is critical for learning about a certain anatomical structure's behaviours in an immersive VR-based environment. This results in responsive stimulus in which the user's input causes an entity in the virtual world to be updated in accordance with the user's actions and/or request. Responsive interactions between a user and object within a virtual environment increases the level of stimulus and interest in learning, which naturally leads to better practical abilities for the user, allowing them to feel more confident through self-directed interactions.

To enable user interactions in the Virtual Anatomy Lab, Unity's XR Interaction Toolkit was imported, allowing 3D and UI interactions via mouse and keyboard. Each interactable component had scripts written in C# implemented to it, defining the object's characteristics and functionality to facilitate interactions between the objects and the user. The toolkit includes simulated Left-Hand and Right-Hand XR Controllers for hand-like interactions, enabling users to hold, pick up, drop, enlarge, or rotate the 3D Skull Model based on step-by-step instructions on tasks displayed in the Task Window UI. This include exercises from medical textbooks presented in textual and graphical information, guiding students to engage with the 3D Skull Model. Such visual feedback enables users to also distinguish between static and interactive objects, which can be identified through colour changes upon hovering of the controllers (see Figure 4).

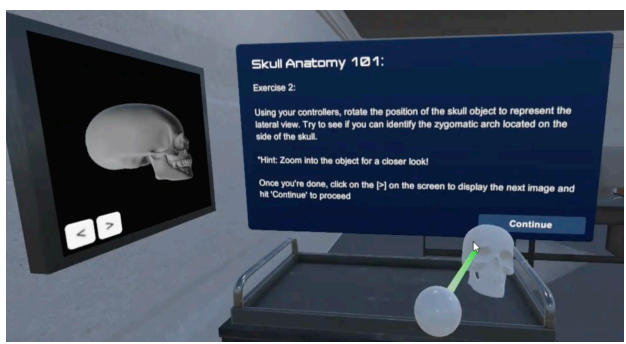

4. Controller Activated for Interaction with 3D Skull Model

These exercises focus on understanding the human skull's anatomy through the five fundamental perspectives: Frontal, Lateral, Superior, Posterior, and Inferior Views, mimicking the practical experience of handling 3D organs in the virtual world as they would with cadavers in a real-life classroom. Overall, the ability for users to control and interact with components in the Virtual Anatomy Lab provides a meaningful learning process, which in turn elevates the student's ability to actively visualise concepts as they are being studied in VR – reducing the mental effort and time that is often required when studying anatomy.

### 3. TESTING AND EVALUATION

The effectiveness of the VR system was essentially evaluated through several testing techniques to verify the quality, performance, and usability of the system. Since the project followed an agile iterative software development approach, testing activities were performed at each stage of the development lifecycle. Its performance was measured by examining the speed and responsiveness of interactions and movements made from the user's point of view while the VR system was running. In case certain components in the virtual environment were found to be lacking in terms of quality or functionality, necessary measures were taken (i.e. redesigning components, refactoring code, implementing additional scripts etc.) to improve the effectiveness and overall user experience. The findings and outcome of testing activities conducted by the intended users of this system are evaluated and explained as follows.

### 3.1. Discussion and Analysis of UAT Results

Fifteen students aged 20 to 22 from five major Malaysian universities participated in the User Acceptance Testing (UAT) which lasted over a duration of two weeks. These participants, active undergraduates from various medical specialties with Human Anatomy in their curriculum, volunteered and consented to participate. UAT sessions, conducted individually and scheduled via Zoom, ranged from 30 minutes to an hour, depending on the user's proficiency with the VR controls and navigation. At the end of each session, participants evaluated the system's usability through a survey.

Taylor's University and International Medical University each accounted for 33.3% of responses respectively, Sunway University 26.7%, and Monash University 6.7%. This diverse participation provided insights into different teaching methods. When asked about study materials, 93.3% answered "Textbook and Coloured Atlas on Human Anatomy" as their primary tool, with "Online Sources" and "Plastic Models" being secondary aids for 86.7% and 80% of respondents, respectively. This aligns with Falah et al. [4] and Triepel et al. [9], who noted the predominance of traditional methods in studying medicine.

Survey responses indicated positive feedback on the VR system's usefulness and convenience. Most 5-point Likert Scale responses ranged between 4 and 5 points. Up to 80% of respondents strongly agreed that learning anatomy through the Virtual Anatomy Lab was more engaging and motivating, which supported the earlier findings from Triepels et al. [9]. Minimal assistance was needed during VR operation, with 73.3% finding navigation easy and 46.7% enjoying interaction with 2D and 3D objects. However, 93.3% remained neutral about the 3D Skull Model's detail and accuracy, suggesting it may not be as comprehensive as expected.

Participants also provided feedback for improvement at the end of the survey. Many suggested adding audio narration for instructions and exercises, and labelling the 3D Skull Model with different colours for structural emphasis. It was believed that the inclusion of audio narration could enhance focus and engagement while interacting with the study materials, replicating an instructor's presence to enhance realism and usability of the VR system.

Although the findings from the UAT were sufficient to support theories and claims made by earlier studies, the study is not without its limitations. Firstly, the UAT was limited to undergraduate students who were actively studying anatomy in the field. Including experienced educators in the testing of the VR system could have provided even more valuable insights, given their extensive experience in teaching the subject. Their feedback could have highlighted how the VR system can be improved further to benefit the overall learning experience for medical students. However, due to various factors, it was challenging to include active teaching professionals from these universities to participate voluntarily. Additionally, the UAT was conducted over a two-week period, limiting the opportunity to recruit a broader pool of participants for the research.

## 4. CONCLUSION

This study developed an immersive and interactive virtual platform to demonstrate the potential of VR as a supplementary tool for enhancing the learning process of human anatomy in medicine. It attempts to address issues and limitations of conventional teaching methods used in medicine by featuring a Virtual Anatomy Lab that was designed to provide a realistic learning experience, encouraging students to explore and experiment with their virtual study materials without ethical concerns or restrictions that may be present in real-life settings. Virtual tasks based on exercises derived from medical textbooks were implemented within the VR system using graphical and textual components to assist students in studying the five fundamental perspectives of human skull anatomy by interacting with the UI components and the 3D Skull Model as the main 3D object for the purpose of the study. These interactions were made possible through the implementation of scripts that contain methods which impose certain behaviours and functionalities to objects and elements in the VR scene. To ensure accessibility, the VR system was built to enable virtual interactions to take place simply through the user's mouse and keyboard inputs, used to provide movement and interactions within the VR scene.

### 4.1. Recommendations for Future Work

To further distinguish this VR system from other comparable VR-based tools used in medical anatomy education, it is strongly recommended that a 'virtual inventory' be added as an extra feature. This aims to widen the field of anatomical research, with the inventory being used to store various 3D organ models to ensure variety and boost customisability of the 3D organ structure examined via VR. Based on the feedback and suggestions provided by the user group in Section 3.1, it is highly suggested for the system to include an audio feature to simulate the user's auditory senses, in addition to the visual simulation within the immersive environment. Additionally, the usability of this system be further improved through the introduction of a haptic-based interaction with 3D object models in the VR environment. This is to widen our scope of research by understanding the potential of creating a tactile and vision-based learning capabilities through the current VR system for medical anatomy education. It is also advisable for future UAT processes to include a number of experienced professionals and educators from the medical field to be to obtain valuable insights for the benefit of the study.

ACKNOWLEDGEMENT

The author of this paper would like to express her sincere gratitude to the students who volunteered to participate in this research.